# Using Survey Data to Obtain More Representative Site Samples for Impact Studies

Robert B. Olsen
George Washington Institute of Public Policy
The George Washington University, Media and Public Affairs Building
805 21st Street NW, Washington DC 20052
(301) 325-1073
RobOlsen@gwu.edu

**Acknowledgements**
The research reported here is supported by the Institute of Education Sciences, U.S. Department of Education, through Grant R305D190020 to Westat.

# Using Survey Data to Obtain More Representative Site Samples for Impact Studies

**Abstract**

To improve the generalizability of impact evaluations, recent research has examined statistical methods for selecting representative samples of sites. However, these methods rely on having rich data on impact moderators for all sites in the target population. This paper offers a new approach to selecting sites for impact studies when rich data on impact moderators are available—but only from a survey based on a representative sample of the impact study's target population. Survey data are used to (1) estimate the proportion of sites in the population with certain characteristics, and (2) set limits on the number of sites with different characteristics that the sample can include. Site recruiters ask potential study sites the survey questions needed to classify them with respect to these characteristics. The Principal Investigator enforces the limits to ensure that certain types of sites are not overrepresented in the final sample. These limits can be layered on top of existing site selection and recruitment approaches to improve the representativeness of the sample.

**Using Survey Data to Obtain More Representative Site Samples for Impact Studies**

  The generalizability of impact studies, particularly randomized controlled trials (RCTs), has become a source of concern in the evaluation field.  This concern is motivated heavily by how the sites are selected for multi-site RCTs.  Participation in most multi-site RCTs is voluntary, and sites recruited to participate in RCTs often choose to opt out. Anticipating this challenge, RCTs often focus their recruitment efforts on sites that are expected to yield the largest samples because their likelihood of agreeing to participate is relatively high, the number of individuals that they could contribute to the sample is relatively large, or both.  This form of site recruitment is often referred to as either purposive sampling or convenience sampling.

  From a design perspective, purposive sampling yields biased impact estimates for the population when the probability of a site's inclusion is correlated with the intervention's impact in that site (Olsen et al., 2013). From a model-based perspective, purposive sampling yields biased impact estimates for the population when the resulting sample differs from the population on characteristics that moderate the impact of the intervention (Tipton, 2013). This bias has been referred to as external validity bias (Olsen et al., 2013). Prior research has shown that purposive sampling for RCTs in education favors districts and schools that are much larger, more urban and more disadvantaged than the average district or school nationwide (Stuart et al., 2017; Tipton et al., 2021). Emerging evidence suggests that purposive sampling can lead to external validity bias:  Bell et al. (2016) found that that the average impact of a federal reading program was significantly smaller in purposive samples of districts than in the population of all districts that received program funding.

In response to these concerns, researchers have developed approaches to selecting samples for impact evaluations or repurposed existing approaches for survey sampling. These approaches fall into one of two categories: (1) design-based sampling, which relies on the site inclusion probabilities determined by the sample design to obtain unbiased estimates, and (2) model-based sampling, which relies on a model of how impacts vary across sites for claims about unbiasedness (e.g., Valliant et al., 2000). Olsen and Orr (2016) proposed stratified random sampling as used in survey research as a design-based option for reducing external validity bias in multi-site impact studies. Tipton (2013) developed a model-based sampling approach that involve k-means clustering to define strata and recruit sites in descending order within strata based on the multivariate distance to the stratum mean. Recent work has tested applications of design-based and model-based approaches just described to two-stage sampling of districts and schools (Litwok et al., under review); it has also introduced and tested more advanced design-based and model-based sampling methods (Fay & Olsen, under review). Both of these studies also tested different methods of replacing sites that decline to participate.

However, there are both logistical and data challenges to implementing the sampling methods listed above. From a logistical perspective, given high-opt out rates, site recruitment teams need to cast a wide net to obtain an adequate sample: Unless the recruiting team starts with a long list of potential sites to recruit, the recruitment process would take too long to complete. However, the sampling approaches referenced tend to focus researchers on smaller sets of sites by rank ordering sites for recruitment in some way—either systematically or randomly—and at least implicitly encourage recruiters to work their way down the list. The focus on "highly ranked" sites from some list can be hard to reconcile with the need to cast the net wide enough to obtain a sample of adequate size.

From a data perspective, all of the design-based and model-based sampling approaches considered thus far benefit heavily from data on site-level moderator variables *for all eligible sites*. Design-based approaches with random selection can use these site-level moderators to stratify the population to improve the precision of the impact estimates; it can also reduce the bias from sites that decline to participate by selecting replacement sites from the same strata as the declining sites. Model-based approaches to sampling sites rely on site-level data on the impact moderators—those in the assumed impact model—for both participating sites and the population of eligible sites to select sites: These data permit model-based methods to select sites that are similar to all eligible sites on factors that are non-ignorable.

However, rich site-level data on impact moderators are rarely available for all eligible sites when an impact study is being designed. Tipton & Olsen (2018) cite several sources of data that can be used to build a population frame and construct impact moderators for populations of districts and schools. These general-purpose data sources tend to include site-level variables of broad general interest: They are not designed to support impact evaluations, much less include data on moderators for every type of intervention that could be rigorously evaluated. Impact evaluations in education often collect richer data on impact moderators over the course of conducting the study. These data are collected specifically because they tend not to be available from existing data sources that can be used to develop a sample design, like NCES's Common Core of Data (CCD).

Rich data on impact moderators are more likely to be available from surveys from nationally representative samples of sites than from population data files that include all eligible sites. For example, the National Teacher and Principal Survey collects data on samples of public schools, private schools, principals, and teachers. The 2017-18 NTBS public school survey included about 10,600 schools from the full population of public schools nationwide as reflected in the CCD (Taie & Goldring, 2019). This survey collects data on a range of variables that may be important moderators for some educational interventions (e.g., variables that capture information on ability groupings, dual enrollment, school start times, and school weeks of less than 5 days). Because these data are collected for a sample of public schools nationwide, they cannot be used to sample public schools from any population of U.S. public schools. At the same time, these surveys offer important information about a broad population of schools that should be useful in designing and conducting impact studies that aim to produce findings which generalize to some pre-defined target population.

This paper proposes and describes a "low-tech" approach to using survey data on potential impact moderators to obtain representative samples of districts and/or schools for impact evaluations of educational intervention. The same general approach could be applied to any substantive field where survey data offers richer data on the population than full population datasets. Under this approach, population estimates from the survey sample are used to set limits on the number of schools with different characteristics that can be included in the study sample. By enforcing these limits, impact study can avoid overrepresentation of certain types of schools in the study sample. This approach can be applied to recruit districts, schools, or both, subject to survey data availability. And it can be used to augment various approaches to sampling sites, from purposive sampling to random sampling. Finally, this approach does not prescribe specific changes to how study teams go about the process of recruiting sites: It simply sets sample limits and allow the research team discretion on how to most efficiently recruit a sample that meets the study's sample size requirements without exceeding those limits.

The next section of the paper describes the four steps required to implement this method. The following section covers its application—both the range of its potential application and a detailed example of how it would be applied. The last section ends with some discussion to put this method in perspective.

**METHODOLOGY**

To implement this method, researchers must: (1) identify a survey that relevant to the impact study in the variables collected and the population covered; (2) estimate population proportions for moderator variables from the survey (or obtain them from published sources); (3) use these survey estimates to set sample size limits for different types of sites in the impact study sample; (4) recruit sites—collecting data on impact moderators from sites interested in participating—while enforcing the sample size limits. These steps are described below.

**Step 1: Identify a Survey that is Relevant to the Impact Study**

To be relevant to the impact survey, the survey must include variables that likely moderate the impact of the intervention and that go beyond the limited variables that are easily available for all schools nationwide (e.g., variables on student demographic characteristics from the CCD). In addition, it must be possible for the research team to obtain or produce population estimates from the survey for a population that aligns closely to the impact study's target population.

When the study's target population is broad in scope—such as all public elementary schools nationwide—published estimates from national surveys may be sufficient to describe the target population. However, when the impact study's target population is a subset of the survey's target population, then either the study team needs published survey estimates for the study's target population *or* access to the survey's microdata needed to produce those estimates. In the latter case, the available microdata needs to include variables that would allow users to isolate or at least approximate the impact study's target population (i.e., variables that can be used to distinguish eligible sites from ineligible sites).

**Step 2: Estimate Population Proportions for Moderator Variables from the Survey**

Using either published survey reports or the survey's microdata, the impact study needs to extract or compute estimates of the share of schools in the impact study's target population with different values of the survey's impact moderators. Suppose that the survey offers $m$ different moderators, and moderator $j$ is a categorical variable with $c_j$ different categories.[1] Then the impact study would need to use the survey to obtain published estimates or calculate estimates of the share of the sites in the study's target population that fall into each of $C = \sum_{j=1}^{m} c_j$ categories. Denote these shares as $p_1, p_2, \ldots, p_C$ and the survey-based estimates of these shares as $\hat{p}_1, \hat{p}_2, \ldots, \hat{p}_C$.

**Step 3: Use Survey Estimates to Set Sample Size Limits for the Impact Study**

---

[1] For this purpose, continuous variables can be converted to categorical variables by setting thresholds (e.g., equals 0 if the variable is below the median and equals 1 if it is above the median).

*Using Survey Data to Obtain More Representative Samples*The survey estimates from the previous step—$\hat{p}_1, \hat{p}_2, \ldots, \hat{p}_C$–serve as targets for the share of sites in the impact study sample that fall into each category. Suppose that *J* is the total number of sites that the study aims to include in the sample, as informed by a statistical power analysis and resource constraints. Then survey estimates indicate that the target number of sites in the study sample from each category should be $J\hat{p}_1, J\hat{p}_2, \ldots, J\hat{p}_C$.

However, these targets may not be integer values. Furthermore, strict adherence to these targets may impede the study's ability to recruit *J* sites. Therefore, the research team can set sample size limits that are somewhat higher than the sample size targets. For example, if the study were willing to accept a sample that differed from the population on the share of sites in any category by no more than 5 percentage points, the study could set the following sample size limits by category: $J(\hat{p}_1 + 0.05), J(\hat{p}_2 + 0.05), \ldots, J(\hat{p}_C + 0.05)$.[2]

**Step 4: Recruit Sites While Enforcing the Sample Size Limits**

In RCTs in education, site recruitment typically occurs over several months (or more). The study team should monitor the recruitment of the sample in real time and keep a running tally of the total number of sites that have agreed to participate in each of the C categories. Critically, once the sample limit in a particular category is reached, the impact study team should not include any additional sites from that category in the sample: Doing so would produce a sample in which sites from that category are overrepresented.

To impose these limits requires that sites be classified based on C categories as they are recruited to determine whether they can be included in the sample. The study team should administer the relevant survey questions to sites as they are recruited to participate, record their responses, and classify them using the C categories. If the site would contribute to sample without exceeding the sample size limit for any of the C categories, the site should be included; if including the site would require exceeding the sample size limit for one or more the C categories, the site should be excluded.

---

[2] Operationally, the impact study team may want round these limits down to the nearest integer to avoid overshooting these limits and exceeding the sample size targets by more than the allowable difference of 5 percentage points.

Once one or more the sample size limits have been reached, the study team face a serious tradeoff between meeting its overall sample size target of J sites, which may benefit from including sites that would exceed one or more sample size limits, and protecting the representative sample by adhering to these limits. To minimize this tension, impact study teams can regularly check their progress toward reaching the sample size targets in each category and reallocate effort to away from categories where they are approaching these targets and toward other categories where additional sites are most needed.

**APPLICATIONS**

With relevant survey data, the methodology described in the previous section can be applied to efforts that recruit districts, schools, or both. Furthermore, the method should be broadly feasible to implement because it doesn't require advance statistical methods or operational changes to the plan for recruiting sites. Finally, in education studies where district cooperation is critical, it doesn't dictate to school districts which particular schools the study should include. Allowing districts to nominate or even select which schools would participate, subject to the constraints of the method described in this paper, may be necessary to obtain cooperation from a sufficient sample of districts and schools.

The remainder of this section provides two examples of how the method could be used in multi-site RCTs of educational interventions.

**Example 1: Hypothetical RCT to Evaluate a K-3 Math Intervention**

Imagine an RCT of a math intervention to be conducted in 40 schools across two states in school districts with which the principal investigators have prior relationships. The impact of this math intervention may depend on a variety of factors that are not available in the CCD or other public databases on all schools nationwide. These factors may include:

1. **The number of hours of math instruction that students receive each week.** For interventions designed to increase the amount of math instruction that students receive, the numbers of hours that students currently receive captures the counterfactual, which likely affects the expected magnitude of the invention's impact.

2. **Whether students stay in the same classes with the same teachers for multiple years**. If the intervention will train teachers in a particular grade level to implement the math intervention, the impact of the intervention in later grades may be larger in schools that keep students with the same teachers because their students may receive a larger dose of the intervention over time.

To implement the method proposed in this paper, the research team could turn to the NTPS, described earlier. The 2017-18 school survey included a question about the number of minutes per day and days per week that third grade students spend on "arithmetic or mathematics" (Section 2, Instructional Time, item 2207). This survey also included a question about whether the student groups [are] assigned to stay in classes together for two or more years with the SAME teacher (i.e., looping)" (Section 3. Students and Classroom Organization, item 2303). The sampling design for the NTPS ensures a sufficient sample size for each state so that "all state-level estimates would meet the criteria for publishability." This suggests that it would be feasible to produce adequately precise survey estimates from the two moderators described earlier for the two states from which the sample will be selected.

To that end, the research team could obtain the survey data under an NCES restricted use license and calculate the following population estimates for the two states together: (1) the thresholds for the second, third, and fourth quartiles (Q2, Q3, and Q4) in the total number of minutes of math instruction per week for third grade students, and (2) the proportion of elementary schools in which students stay with the same teachers for multiple years (p). These estimates would be used to set six sample size targets. For the minutes of math instruction for third grade students, the study should aim to include one-quarter of the total schools in the sample—or 10 schools—from each quartile (four sample size targets). In addition, for staying with the same teacher, the study should aim to include 40p schools in which students stay with the same teachers in multiple years and 40(1-p) schools in which students do not stay with the same teachers in multiple years (two sample size targets). If, for example, students stay with the same teachers in 16 percent of elementary schools in these two states, the study would aim to include 6.4 schools in which students stay with the same teachers in multiple years and 33.6 schools in which students do not. Finally, to set more forgiving sample size limits, the study could set sample size limits that are a little higher than these targets (e.g., up to 11 or 12 school from any one quartile).

**Example 2: A Hypothetical National RCT of a Reading Intervention for English Learners**

Imagine an RCT of a reading intervention for English learners (Els), and suppose that the RCT will require a sample of 80 schools across the country. The study plan involves recruiting approximately 10 school districts and negotiating with them to include a total of 80 schools in those districts in the RCT.

The impact of this intervention may depend on a variety of factors that are not available for all eligible schools, including (1) whether the school offers reading instruction to ELs in regular English-speaking classrooms and (2) whether the school uses ESL, bilingual, or immersion techniques in teaching these students English. Both factors would plausibly influence the effectiveness of classroom interventions for ELs; both are captured by the NTPS 2017-18 school survey (Section 5. Special Programs, items 424 and 425).

For example, suppose that 82 percent of schools teach some ELs using ESL, bilingual, or immersion techniques (so 18 percent do not), and 37 percent of schools teach some ELs in regular English-speaking classrooms (so 63 percent do not). The study should set targets of including: (1) 65.6 schools that teach some English learners using ESL, bilingual, or immersion techniques, (2) 14.4 schools that do not teach any English learners using ESL, bilingual, or immersion techniques, (3) 29.6 schools that teach some English learners in regularly English-speaking classrooms and (4) 50.4 schools that do not teach any English learners in regularly English-speaking classrooms. For each target, the researchers could set their limits slightly higher than their targets to permit small differences between the final sample of schools and the national population of schools nationwide.

However, before recruiting schools, the study team would need to recruit and negotiate with school districts. Suppose that the study team was unable to find relevant survey data on

districts—a plausible scenario since more surveys in education focus on schools than on districts. In this case, the impact study could rely on the CCD to provide some basic district-level information on factors that might moderate the impact of the intervention and could be used to stratify eligible districts for balanced or probability sampling. For example, to ensure that the typical district selected has at least 8 schools that could participate—necessary to obtain 80 schools in only 10 districts—the study could select districts proportional to the number of schools and recruit 8 schools per district. More generally, researchers would have all of the standard sampling methods and population data at their disposal for selecting districts for recruitment.

**DISCUSSION**

The method described here is not a sampling method or design: it is simply a tool to ensuring that site recruitment efforts avoid producing samples of sites that differ substantially from the populations about which impact studies hope to learn. This tool could allow researchers to control the final sample with respect to richer sets of potential moderators than most sampling methods, which rely on assembling moderator data on all eligible sites. As such, it can be thought of as a "model-based enhancement" to whatever sampling method is used— and if no formal sampling method is used, a way of constraining the composition of a purposive or convenience sample.

Unlike probability sampling with large samples and universal participation, the method described here cannot guarantee that the sample closely matches the population on *unobserved* impact moderators. But participation by recruited sites is far from universal in most RCTs. And with high opt-out rates by sites, no sampling method is sufficient to support claims of generalizability without some additional assumptions about which variables moderate the impact of the intervention. Put differently, when opt-out rates by recruited sites is high, the assumptions required by pure design-based sampling methods will not hold. In this setting, claims of generalizability necessarily rest on the balance between the sample and the population.  The method described here is designed to help researchers obtain samples that are comparable to the population on a richer set of potential moderators to increase the plausibility that the impact estimates from the sample generalize to the population.

    Finally, a key virtue of this proposed method is feasibility of implementation. It does not require advanced statistical skills, high-cost data collection efforts, or disruptive changes to standard practice in recruiting sites.  The method can be implemented by any research team capable of conducting an impact evaluation:  It requires the ability to produce descriptive statistics from standard surveys and to do simple calculations of the type shown in the applications section.  Given the feasibility of this low-tech method, it may help bridge the wide gap between the current literature on generalizability, which focuses on advanced statistical methods, and the usual practice of recruiting sites for multi-site RCTs.